\documentclass[conference, a4paper]{IEEEtran}
\IEEEoverridecommandlockouts
\usepackage{cite}
\usepackage{amsmath,amssymb,amsfonts}
\usepackage{graphicx}
\usepackage{textcomp}
\usepackage{xcolor}
\usepackage{algorithm}
\usepackage{algpseudocode}
\usepackage{mathtools}
\usepackage{url}

\def\BibTeX{{\rm B\kern-.05em{\sc i\kern-.025em b}\kern-.08em
    T\kern-.1667em\lower.7ex\hbox{E}\kern-.125emX}}
\begin{document}
\addtolength{\headsep}{0.1cm}
\title{Multi-Day Scheduling for Electric Vehicle Routing: A Novel Model and Comparison Of Metaheuristics\\
\thanks{We would like to thank R. Schoenen and Bertrandt Technologie GmbH Regensburg for their support and the opportunity to participate in the GUIDE project, which formed the basis for this study (cf. www.bertrandt.com/software-experts/bertrandt-guide). We gratefully acknowledge the support received from the Regensburg Center of Energy and Resources (RCER) (cf. www.rcer.de)}
}

\author{

\IEEEauthorblockN{Dominik Köster}
\IEEEauthorblockA{\textit{Faculty of Computer Science and} \\
\textit{Mathematics, OTH Regensburg}\\
Regensburg, Germany \\
dominik.koester@oth-regensburg.de}
\and
\IEEEauthorblockN{Florian A. Porkert}
\IEEEauthorblockA{
\textit{Bertrandt Technologie GmbH}\\
Regensburg, Germany \\
florian.porkert@bertrandt.com}
\and
\IEEEauthorblockN{Klaus Volbert}
\IEEEauthorblockA{\textit{Faculty of Computer Science and} \\
\textit{Mathematics, OTH Regensburg}\\
Regensburg, Germany \\
klaus.volbert@oth-regensburg.de}
}

\maketitle

\begin{abstract}
The increasing use of electric vehicles (EVs) requires efficient route planning solutions that take into account the limited range of EVs and the associated charging times, as well as the different types of charging stations.
In this work, we model and solve an electric vehicle routing problem (EVRP) designed for a cross-platform navigation system for individual transport. 
The aim is to provide users with an efficient route for their daily appointments and to reduce possible inconveniences caused by charging their EV.
Based on these assumptions, we propose a multi-day model in the form of a mixed integer programming (MIP) problem that takes into account the vehicle's battery capacity and the time windows of user's appointments.

The model is solved using various established metaheuristics, including tabu search (TS), adaptive large neighborhood search (ALNS), and ant colony optimization (ACO).
Furthermore, the performance of the individual approaches is analyzed using generated ensembles to estimate their behavior in reality and is compared with the exact results of the Google OR-Tools solver.
\end{abstract}


\section{Introduction}\label{sec:Intro}
In 1959 Dantzig and Ramser introduced the vehicle routing problem (VRP) \cite{DR_1959}, a generalization of the traveling salesman problem that has since spawned numerous variants and practical applications in logistics and transportation \cite{Toth.2014, Erdogan.2012, gendreau2010handbook, ReviewOfReviews, Asghari, Qin, Xiao, Kucukoglu}. The increasing adoption of electric vehicles (EVs) motivates a specialized class of these problems, the electric vehicle routing problem (EVRP), which extends the VRP by integrating battery limitations and charging requirements.

Beyond the standard VRP aspects, EVRP formulations must explicitly model battery range and available charging options, since recharging can introduce detours and delays that materially affect routing decisions \cite{Qin}. These additions introduce many new parameters for practical models, and due to the combinatorial problem structure, lead to rapidly growing solution times as instance size increases \cite{Asghari, Ye.2022}. Addressing these challenges is essential for scalable solutions in smart mobility and vehicular applications.

This work focuses on efficiently scheduling a typical user’s multi-day appointments under EV constraints, with the ultimate goal of extending the model to fleets. While the current formulation considers a single vehicle, formally an EV traveling salesman variant, we treat it as an EVRP with fleet size of one to align with the broader literature. We propose a mixed-integer programming (MIP) model for time-fixed and time-flexible events that allows for parallel charging during appointments and supports tunable objective weights for time, distance, and end-of-route state-of-charge. Finally, we compare the OR-Tools MIP solver \cite{ortools} with three metaheuristics: Tabu search (TS), adaptive large neighborhood search (ALNS), and ant colony optimization (ACO). Our implementations have been integrated into automotive software and evaluated in an industrial setting to demonstrate practical relevance for smart cities and vehicular technology. In summary, we (i) introduce a multi-day EVRP formulation supporting parallel charging and time-flexible events, (ii) evaluate exact methods and metaheuristics on randomized instances, and (iii) propose a practical hybrid strategy to balance solution quality and runtime for different problem scales.

\section{Related work}\label{sec:rel}
In recent years, considerable research effort has been invested in the solution strategies of vehicle routing problems and, in particular, EVRPs, which have led to an enormous amount of relevant literature (cf. \cite{ReviewOfReviews}).
In fact, some problems seem to be quite amenable using exact methods, especially some of the methods included in the leading commercial MIP software packages, for example, those related to IBM's CPLEX \cite{cplex12} or Google OR-Tools \cite{ortools}.

However, determining the optimal solution for EVRPs is NP-hard \cite{Erdogan.2012}, so commercial applications therefore tend to use more flexible strategies such as metaheuristics due to the size and frequency of the real-world EVRPs they are designed to solve. The underlying approach can be roughly divided into exact, heuristic, metaheuristic, and hybrid methods (cf. \cite{Ye.2022, Xiao, Erdogan.2012}). Reference \cite{Moghdani} found that $34\%$ of the methods used in related research were exact methods, while $41\%$ of the problems were solved using metaheuristics.

Moreover, the characteristics of an EVRP depend strongly on the optimization objectives. To model the complexity of EV charging in MIP formulations, common approaches include duplicating charging station nodes \cite{Erdogan.2012}. Moreover, EVRP with charging (EVRP-C) can involve strategies of partial \cite{Xiao} or full charging \cite{Erdogan.2012}, and battery models may be linear \cite{Schiffer.2017} or nonlinear, with correlated charging rates \cite{Xiao}.  \\

In addition to vehicle energy consumption, parameters such as charging time and fleet size play a key role in realistic EVRPs \cite{ReviewOfReviews}. EVRP with variable time windows (EVRP-TW) considers limited charging times between or parallel to specific time windows \cite{Schneider.2014}.
In particular, when considering fixed time windows, an intermediate charging stop must not overlap with another specified time window; otherwise, a solution would be invalid. EVRPs with a fixed fleet size (EVRP-MF) were first introduced by \cite{Gorgi}. Later, \cite{HIERMANN2016995} extended this scenario to include the use of publicly accessible charging stations \cite{Xiao}. 

To our knowledge, there is currently no detailed model that takes into account parallel charging at events, multi-day planning, and the minimization of route distance and travel time.

\section{Problem description}\label{sec:prblemDescription}

An optimal route is computed from the user's appointments, also called events, which are either \textit{time-fixed} (fixed start and end) or \textit{time-flexible} (only a duration). Event locations are either explicit coordinates or predefined types (e.g., \textit{Groceries}, \textit{Parking}, \textit{Restaurant}). For these locations, a limited set of candidate locations is obtained from map provider APIs. For $N$ flexible events $\prod_{i\in N}\lambda_i$ route combinations must be considered, where $\lambda_i$ is the number of locations for event $i$.

We support multi-day planning via so called \textit{separator events} that split days by specifying a latest arrival, earliest departure and a minimum duration of the event (e.g., home 19:00–07:00). Separator events are fixed in the event order but time-flexible regarding earlier arrival, and any saved time is reflected in the objective function.

Charging stations are retrieved from a charging station API using event coordinates and a maximum walking distance. We preselect one charging station per event using a charging score that accounts for charging power, walking distance, and number of charging plugs, and schedule charging preferably in parallel with these events. Additional charging stations are provided to users as backups in case the selected station is unavailable. This factor is considered, and an extra buffer is added to the walking distance to allow for another charging station. 

Similar to charging stations and event locations, we retrieve distances between these points using an API that specializes in real-time traffic data. Using this type of provider enables us to model the optimization problem based on real-world conditions. In the case of charging stations, for example, data about charging station occupancy makes applications based on the solution to this problem very practical in the real world.

In addition, the user specifies a preference among the shortest distance, the shortest travel time, and the highest possible battery level at the end of the route. General battery information is sent to us directly from our vehicle, including the charging plug type, the maximum supported charging power, and the remaining vehicle range. This remaining range is also the value used by the mathematical model when calculating the cost of edge traversal.

The corresponding EVRP is defined on a complete directed graph $G =(V_0,E)$, where $V_0$ is the vertex set including the start node. $E$ is a set of edges connecting the vertices $V_0$. All symbols for the following mathematical model are listed in Table \ref{tab:model_symbols}. The goal of optimization is to find the shortest route for the vehicle, end each day as early as possible, maintain enough battery charge to safely traverse the route, maximize battery capacity at the end of the route, and minimize charging stops during the trip.

\begin{table*}
\caption{Symbols of the EVRP model}
{\begin{tabular}{@{}p{0.47\textwidth}l@{}p{0.45\textwidth}c@{}}
\hline
\hspace{0.8pt}
Symbol definitions                                                                  & Symbol definitions \\ 
\hline 
$V_0$ Index set of events represented by nodes, $V_0\!=\!\{ 0, \, \ldots \,, n-1\}$,& $a_{u,\min}$ Earliest arrival time at node $u$, $u \in V_0$ \\
\quad where $0$ is the start node and $n-1$ is the end node                         & $a_{u,\max}$ Latest departure time at node $u$, $u \in V_0$\\
$V$ Set of events represented by all nodes excluding the start node,                & $k^{\min}$ Minimum range of the battery \\
\quad $V\!=\!V_0 \setminus \{0\}\!=\!\{ 1, \, ...\,, n-1\}$                         & $k^{\max}$ Maximum range of the battery\\
$E$ Set of edges between two nodes, $E\!=\!\{(u,v)\;|\;u,v\in V_0\}$                & $T_u^L$ Time for walking distance between node $u$ and the corresponding \\
$S$ Set of separator nodes between days                                             & \quad charging station \\
$d_{u,v}$ Distance between node $u$ and $v$, $(u,v) \in E$                          & $c_u$ Maximum potential charge gain at node $u$, $u \in V_0\setminus \{n-1\}$\\
$t_{u,v}$ Travel time from $u$ to $v$, $(u,v) \in E$                                & $a_{u}$ Arrival time at node $u$, $u \in V_0$ \\
$A_{u}$ Constant arrival time at node $u$, $u \in V_0$                              & $k_u$ Battery range on arrival at node $u$, $u \in V_0$ \\
$a_{0}$ Start time at the first node $0$, $0 \in V_0$                               & $x_{u,v} \, 0-1$ variable, $1$ when the edge between nodes $u$ and $v$ is used, $0$  \\
$D_{u}$ Duration of stay at node $u$, $u \in V_0$                                   &  \quad otherwise, $(u,v) \in E$\\
$w \in [0,1]^3$ Weighting of distance, time and charging optimization               &  $r_u$ $0-1$ variable, $1$ when node $u$ is used for parallel charging, $0$ otherwise,\\
\quad in the objective function, $w = (w_d,w_t,w_c)$                                & \quad  $u \in V_0\setminus \{n-1\}$  \\
$\varepsilon \in [0,1]$ Weighting for the total amount of charging stops and the    &  $\tilde{c}_u$ Range gain by charging at node $u$, $u \in V$\\
\quad depature time at $a_0$                                                        & \\\hline
\end{tabular}}{}
\label{tab:model_symbols}
\end{table*}

The novel mathematical model for the EVRP problem, introduced in this work, is formulated as follows:\\
\begin{align}
        \min\quad w_d\sum_{(u,v) \in E} x_{u,v} d_{u,v} + w_t \sum_{u \in S} (a_{u} - s(u)) \nonumber \\ 
        \quad - w_ck_{n-1} + \varepsilon \sum_{i \in V_0 \setminus \{n-1\}} r_i + (\varepsilon a_{0})
\label{eq:objective_function}
\end{align}
s.t.
\begingroup
\allowdisplaybreaks
\begin{align}
        \label{eq:Flussbedinung_In}
        &\sum_{v \in V_0} x_{u,v} = 
        \begin{cases} 
            1, & u \in V_0 \setminus \{n-1\},\\ 
            0, & u = n-1 ,
        \end{cases}\\ 
        \label{eq:Flussbedinung_Out}
        &\sum_{v \in V_0} x_{v,u} = 
        \begin{cases} 
            1, & u \in V\setminus\{0\},\\ 
            0, & u = 0,
        \end{cases}\\
        \label{eq:Eigenzyklen}
        &x_{u,u} = 0, \forall u\in V_0, \\
        \label{eq:Zyklen}
        &x_{u,v} + x_{v,u} \leq 1, \forall u,v\in V_0, \\
        \label{eq:KonsAnkunftszeiten}
        &a_u = A_u - r_uT_u^L, \forall u \in V_0,\\
        &a_{u,\min} - (r_uT_u^L) \leq a_{u} \leq a_{u,\max} - D_u - (r_uT_u^L), \nonumber\\
        & \quad\forall u \in V_0 \setminus S, \nonumber\\
    \label{eq:WertebereichAnkunftzeiten}
	&s(u) + (r_u T_u^L) \leq a_{u} \leq a_{u,\max} - D_u - (r_u T_u^L), \nonumber\\
    &\quad \forall u \in S, \\
    &a_u + D_u + t_{u,v} + 2(r_u T_u^L) < a_v - (r_vT_v^L), \nonumber\\
    &\quad \forall u,v \in V_0 \setminus \{u \in S\} \text{ with } x_{u,v} = 1, \nonumber \\
    \label{eq:ZusammenhangAnkunftszeiten}
    &a_{u,\max} + t_{u,v} + (r_u T_u^L) < a_v - (r_v T_v^L), \forall u \in S, \nonumber\\
    &\quad \forall v \in V_0 \text{ with } x_{u,v} = 1, \\
    \label{eq:LadenAmEndknoten}
    &r_{n-1} \coloneqq 0, \\
    \label{eq:Mindestreichweite}
    &k_{\min} \leq k_u, \forall u \in V_0, \\
    \label{eq:Reichweitenzugewinn}
    &\tilde{c}_u \leq r_u c_u, \forall u \in V_0 \setminus \{n-1\}, \\
    \label{eq:MaximalerAkku}
    &k_u + \tilde{c}_u \leq k_{\max}, \forall u \in V_0 \setminus \{n-1\}, \\
    \label{eq:Ankunftsreichweiten}
    &k_u + \tilde{c}_u -d_{u,v} = k_v, \forall u,v \in V_0  \text{ with } x_{u,v} = 1, \\
    \label{eq:a}
    &a_u \in \mathbb{R}_0^{+}, \forall u \in V_0, \\
    &k_u \in \mathbb{R}_0^{+}, \forall u \in V_0, \\
    &x_{u,v} \in \{0,1\}, \forall (u, v) \in E, \\
    \label{eq:r}
    &r_u \in \{0,1\}, \forall u \in V_0 \setminus \{n-1\}.
\end{align}
\endgroup

Where \eqref{eq:objective_function} is the objective function of the total travel distance, the total travel time, the charging level at the end of the route respectively weighted with their constants $w$. Additionally, there are two summands that are scaled with a small $\varepsilon$, the number of charging stations and the earliest departure time. The $\varepsilon$ summands are there to give the objective function an incentive to minimize the number of charging stations and to start the route as early as possible. The individual constants for the target function $w$, are composed of a user preference and a normalization of the respective solution summand. User preferences for $w_d$, $w_t$, and $w_c$, are in sum $1$ and given by the user. The constants $w$ are then normalized by determining the upper and lower limits of a solution summand. For example, the total travel distance is a percentage of the distance between the best possible solution and a suboptimal one. To achieve this, a simple initial solution is first found using greedy heuristics. This solution is used for an upper suboptimal bound of the objective function. In our case, by using the Best-Fit-Decreasing (BFD) heuristic, which sorts and inserts the time-flexible nodes between time-fixed nodes similar to a bin-packing problem \cite{BinPacking}. To estimate a lower bound for normalization, the shortest edge of each node is added. More precise methods exist, such as using a minimum spanning tree with Prim's or Kruskal's algorithm, as mentioned in \cite{cormen2022introduction}. The aim of this procedure is to normalize the relationship between the summands of the objective function to ensure they all have the same influence on it. The more precisely the limits of the respective summands can be estimated, the more precisely the target function summands are in the same relationship to each other \cite{MIP}.

Constraints \eqref{eq:Flussbedinung_In} and \eqref{eq:Flussbedinung_Out} denote that every node, except the start and end node, has exactly one incoming and one outgoing edge. Constraint \eqref{eq:Eigenzyklen} and \eqref{eq:Zyklen} forbid self-cycles and cycles between two nodes. Constraint \eqref{eq:KonsAnkunftszeiten} determines that if a constant arrival time $A_u\!<\!\infty$ is defined for a node, it must be taken into account. Constraints \eqref{eq:WertebereichAnkunftzeiten} indicate that the actual arrival time at a node must lie between the earliest arrival time and the latest departure time minus the time spent there. In the case of a separator node, the earliest departure time at a node is the latest departure time of the previous separator node. If a charging process is planned in parallel at a node, the walking distance between the charging station and the node must be included in the arrival and departure times. Constraints \eqref{eq:ZusammenhangAnkunftszeiten} specify that if an edge is selected (e.g., $x_{u,v} = 1$), the arrival time of the destination node must be greater than the departure time of the start node and the traversal time between the two. The departure time is made up of the arrival time, the duration at the node, and the possible walk to the charging station. If $u$ is a separator node, the clearly defined latest departure time of the node is used for this purpose. Constraint \eqref{eq:LadenAmEndknoten} forbids charging at the end node. Constraint \eqref{eq:Mindestreichweite} represents the minimum battery capacity that a vehicle should reserve. Constraints \eqref{eq:Reichweitenzugewinn} and \eqref{eq:MaximalerAkku} specify the actual charge gain at a node and that it cannot exceed the maximum capacity of the battery. Constraint \eqref{eq:Ankunftsreichweiten} states that when selecting an edge between two nodes (e.g., $x_{u,v} = 1$), the arrival range at a node is precisely the battery range of the previous node, including any additional charge gained through charging, minus the cost of the corresponding edge. Variables \eqref{eq:a} to \eqref{eq:r} define the solution variables that need to be solved in order to receive a valid solution.

For the objective function and for the restriction of the arrival times, the function $s(u)$ is used for the separator nodes, which is defined as follows:
\begin{equation}
    \label{eq:AnkunftszeitpunkteFunktion}
    \begin{aligned}
        s(u) = 
        \begin{dcases*} 
            a_0, & if $u$ is the first element in set $S$,\\ 
            a_{v,\max}, & otherwise, where $v$ is the \\
            & predecessor element of $u$ in $S$.
        \end{dcases*} 
    \end{aligned}
\end{equation}

The listed constraints are shown here in their basic simple form and have to be linearized for use in a MIP solver (e.g. constraints \eqref{eq:ZusammenhangAnkunftszeiten} and \eqref{eq:Ankunftsreichweiten}). For this purpose the \textit{Big-M-Method} is typically used to achieve multiple linear equations \cite{MIP}.
\section{Problem-solving method}\label{sec:AI}

\subsection{Exact solver} \label{sec:exact_solving}
The first approach consists of the MIP solver from Google OR-Tools \cite{ortools}. The solver we use is the \textit{SCIP} solver, which is used by default by the OR-Tools. The mathematical model is set up with the help of auxiliary functions that generate solver variables and constraints. An increased number of variables and constraints leads to an increase in the complexity of the MIP, which in turn is reflected in a longer solution time by the solver. More variables increase the search space exponentially, while additional constraints can increase the solution time by making the relaxation problem of the linear system of equations more complex \cite[Chapter~4]{MIP}. 

\subsection{Tabu Search (TS)} \label{sec:TS}
As introduced by \cite{Glover}, TS is an iterative, single-solution metaheuristic that balances local search and strategic diversification via a short-term tabu memory. The TS framework discussed in this paper follows the approach of \cite{Wang}. First, the BFD heuristic is used to find an initial solution. Then, in each iteration, the neighborhood of the solution is explored, and the best non-tabu solution is sought. In this context, a solution is a list of event nodes, and neighborhoods are defined by elementary moves such as swapping or inserting these nodes. A swap operation involves swapping two nodes in the solution, while an insertion involves placing a node between two other nodes. In our case, one favorable restriction for this method is that a valid solution must maintain the total order of the time-fixed events: for the list of events $( E_1, E_2, \ldots, E_n )$ for all $u \in V_0$ with $A_u\!<\!\infty$ applies:
\begin{align*}
  \forall i, j \in \{1, 2, \ldots, n\}, \; i < j \implies t^{end}_i < t^{start}_j.
\end{align*}  
Two discrete tabu lists are maintained to track these distinct move types and prevent immediate reversals. After each candidate move, a greedy charging planner evaluates the route's feasibility and cost.

For this, candidate charging stations along the route are ranked according to the potential battery gain relative to time loss. Stops are added until the route can be completed safely and no further addition improves the objective function \eqref{eq:objective_function}. In some cases an additional stop is justified if the incremental charge exceeds a threshold (e.g. $10\%$ of total capacity). In the event of $w_c = 0$, it follows that only the minimum number of stops is scheduled. The search procedure incorporates the selection of the optimal non-tabu neighbour, with the process terminated after a predetermined number of iterations. The lengths of tabu lists remain a pivotal tuning parameter, governing the trade-off between intensification and diversification.

\subsection{Adaptive Large Neighborhood Search (ALNS)} \label{Lösungsansätze_ALNS}
ALNS is a single-solution, randomized metaheuristic that explores large neighborhoods by iteratively destroying and repairing portions of the current solution \cite{Pisinger/Ropke_ALNS}. As in TS, ALNS starts from a BFD initial solution. In each iteration, a destruction operator is applied to remove a subset of nodes, and a repair operator is subsequently implemented to reconstruct a feasible route. The resulting candidate is then evaluated by the objective function \eqref{eq:objective_function}, which determines future operator selection. The selection of operators is a probabilistic and adaptive process. Operators that have demonstrated a propensity for producing favorable outcomes in prior iterations are selected with a higher probability, thereby promoting an empirically effective search balance.

Destruction is defined as the removal of a configurable number of nodes. This number is referred to as the degree of destruction (DoD). In this study, we examine three schemes: a fixed DoD, an increasing DoD that grows over iterations and may eventually remove the full solution, and a randomized DoD sampled per iteration between one node and the entire solution. These options trade exploration and exploitation, while small DoDs favor local refinement, and large DoDs enable global diversification.

The repair process utilizes one of three operators to restore feasibility. The initial procedure entails random repair, which involves the reinsertion of nodes at random positions while assessing the feasibility within a designated time window. Constructive repair, on the other hand, involves the repeated insertion of the node that yields the smallest incremental cost with respect to time and distance from the set that has been removed. Furthermore, an exact MIP-based repair \cite{ALNS_MIP} is incorporated for moderate-size subproblems. A subset of decision variables is fixed to the partial solution, and the solver completes the remainder exactly. The performance of each repaired route is evaluated using the same greedy charging planner employed by TS. The combination of heuristic and exact repairs, in conjunction with adaptive selection, ensures robustness across instance sizes while maintaining per-iteration effort tunability.

\subsection{Ant Colony Optimization (ACO)}
In this work, ACO is adapted from \cite{Dorigo2010_ACO} and constructs multiple ant routes by stochastic node selection guided by pheromone values, denoted by $\tau_{ij}$, and heuristic information, denoted by $\eta_{ijk}$. The probability of selecting node $j$ (when moving from node $i$ towards $k$) is given by 
\begin{equation}
    \label{eq:P}
    \begin{aligned}
        P_{ijk} = \frac{[\tau_{ij}]^\alpha \cdot [\eta_{ijk}]^\beta}{\sum_{e \in G}[\tau_{iek}]^\alpha \cdot [\eta_{iek}]^\beta},
    \end{aligned}
\end{equation}
where $\eta_{ijk}$ encodes heuristic desirability (in this case, the inverse of the combined distance/time costs) and $\alpha$ and $\beta$ control the relative influence of pheromone and heuristic information. Each ant constructs an entire route utilizing this distribution. The route is then evaluated after applying the same greedy charging planner used for the other metaheuristics.

Subsequent to the construction of all ant-solutions, pheromones are updated to influence future search. We adopt the standard evaporation and reinforcement rule \cite{Dorigo2010_ACO}:
\begin{equation}
\tau_{ij} = (1 - \rho) \cdot \tau_{ij} + \sum_{s \in \{S_{iter} \cup s_{best}\}} g(s),
\end{equation} 
where $\rho$ is the evaporation rate, and $g(s)$ is an evaluation term. For the edges used in solution $s$, we employ $g(s)$ to be equivalent to $1/f(s)$, with $f(s)$ being the objective function from \eqref{eq:objective_function}. Evaporation prevents premature convergence by reducing legacy pheromone influence, while reinforcement increases pheromone on edges that appear in high-quality solutions.

\section{Results} \label{sec:Results}
All experiments were implemented in C\# (.NET) using Visual Studio 2022 and executed on an Intel® Core™ i5-1145G7 at 2.60 GHz with 16 GB RAM. For each test case, we generated 100 random EVRP instances. The instance generation process queried and stored API-dependent data in advance. Each instance generated a random number of days and up to 120 events. For parallel charging during events, the API returned nearby stations for each event. One station per event was selected based on a simple score combining distance, station power, and the number of compatible plugs. All locations, coordinates, stations, and pairwise distances were fixed in the problem formulation so that API latency would not affect runtime measurements. In the following experiments, the quality of a solution refers to the exact solution computed with the MIP solver and unlimited runtime.

\begin{figure} [t]
    \centering
    \includegraphics[width=0.85\columnwidth]{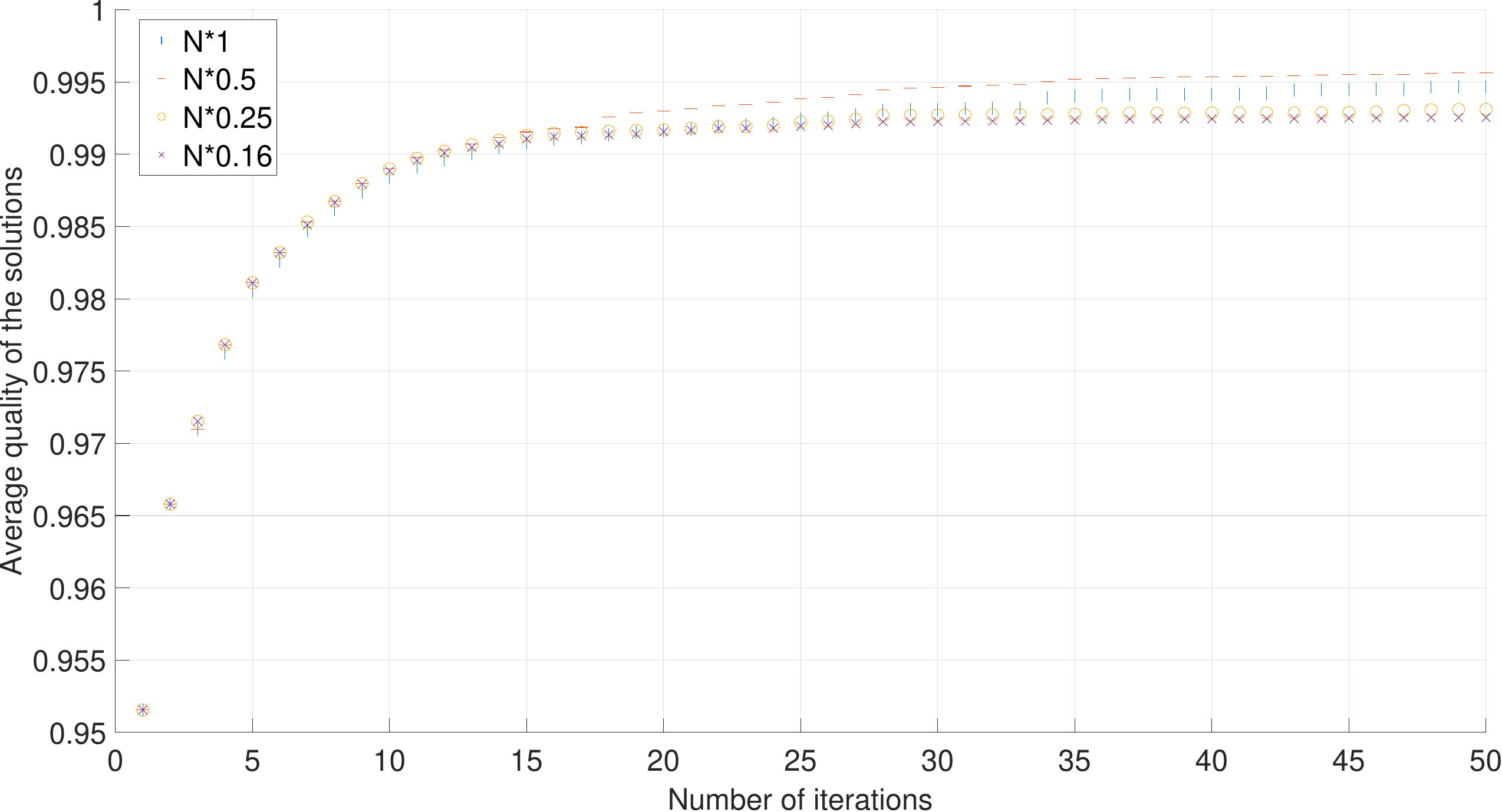}
    \caption{The average quality of TS for tabulists of different sizes.}
    \label{fig:tabulists}
\end{figure}

\begin{figure} [t]
    \centering
    \includegraphics[width=0.85\columnwidth]{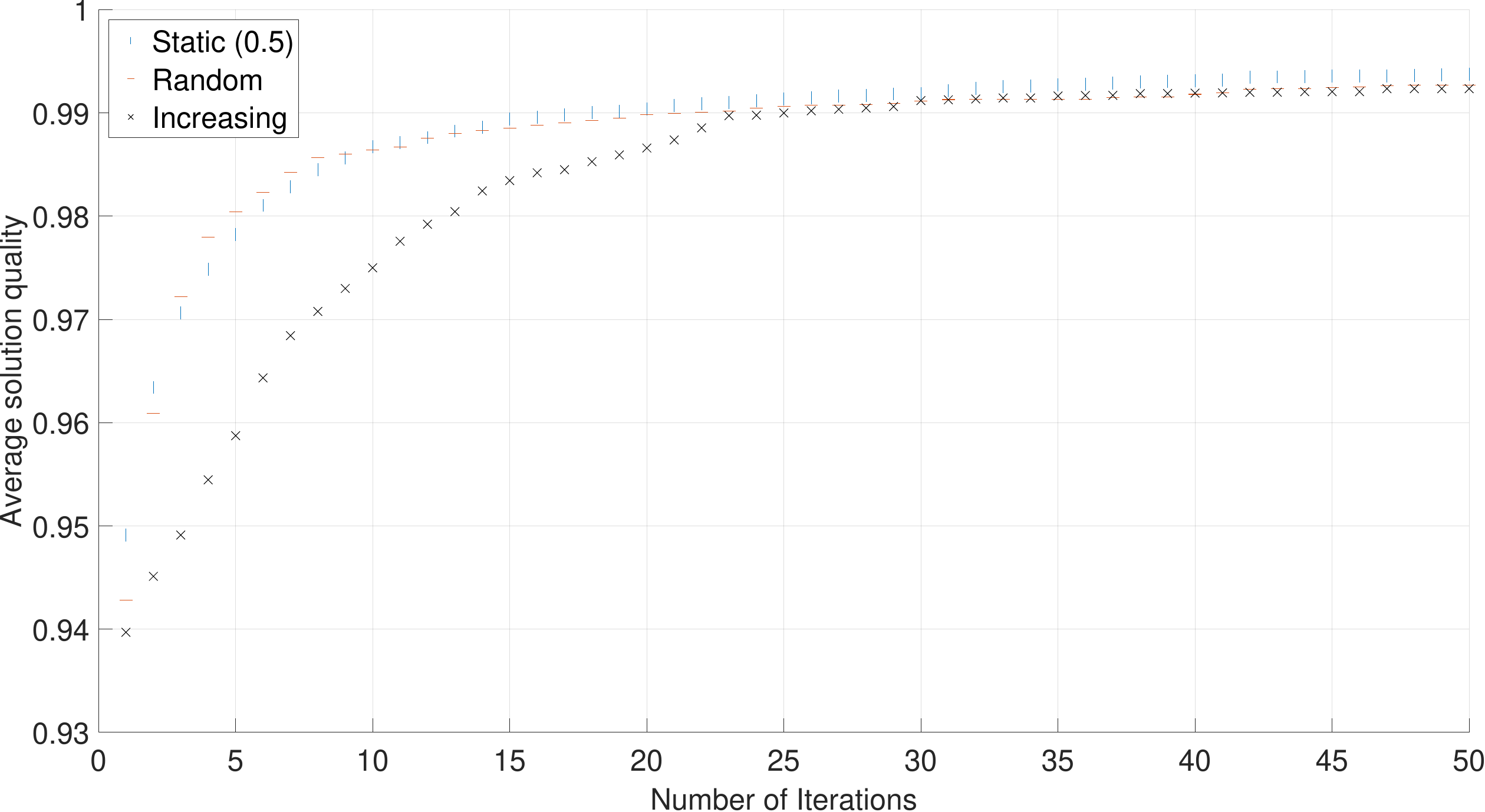}
    \caption{The average quality in each iteration with different DoDs.}
    \label{fig:repairResults}
\end{figure}

For TS a key parameter is tabu-list size. We tested commonly suggested values ($N/6$, $N/4$, $N/2$, $N$, \cite{Gmira}) where $N$ denotes the number of events. Results are shown in Fig. \ref{fig:tabulists}. Initially the lists are empty and methods perform similarly, but after warm-up $N/2$ yields the best average solution quality, followed by $N$.

\begin{figure} [t]
    \centering
    \includegraphics[width=0.85\columnwidth]{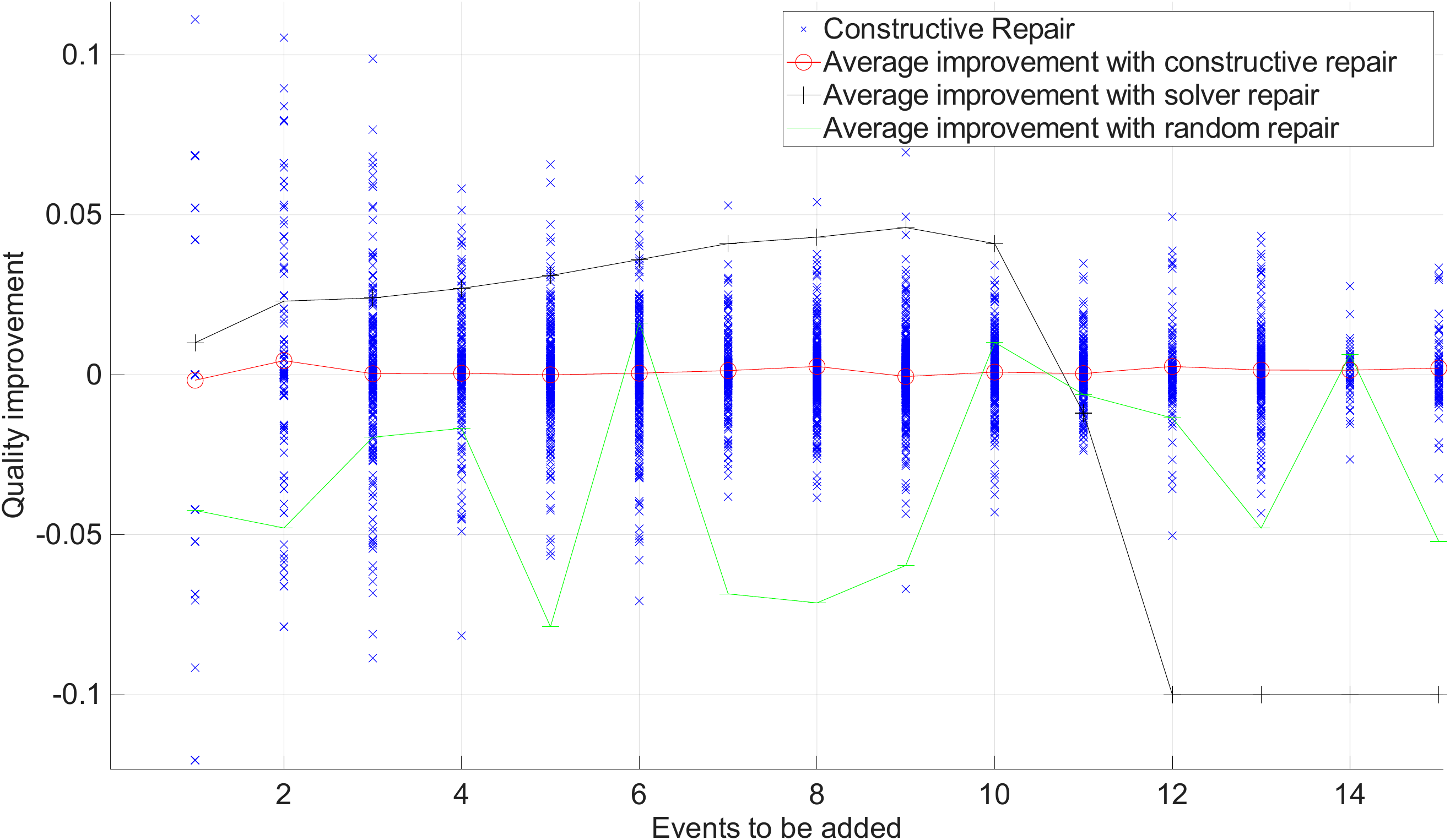}
    \caption{The improvements in quality with different repair operators.}
    \label{fig:alnsResults}
\end{figure}

For ALNS both destruction and repair choices matter. Fig. \ref{fig:repairResults} compares three destruction schemes. A static DoD of 0.5 produced the best average quality. For repair (Fig. \ref{fig:alnsResults}), constructive repair is sometimes beneficial but not consistently so, random repair performs worse on average, and the exact MIP-based repair yields a significant improvement up to a certain subproblem size. If the MIP solver exceeds the per-test cutoff it is recorded as a timeout (plotted at -0.1), indicating the solver fails to improve within the allotted time from roughly 10 removed events onward.

\begin{figure} [t]
    \centering
    \includegraphics[width=0.85\columnwidth]{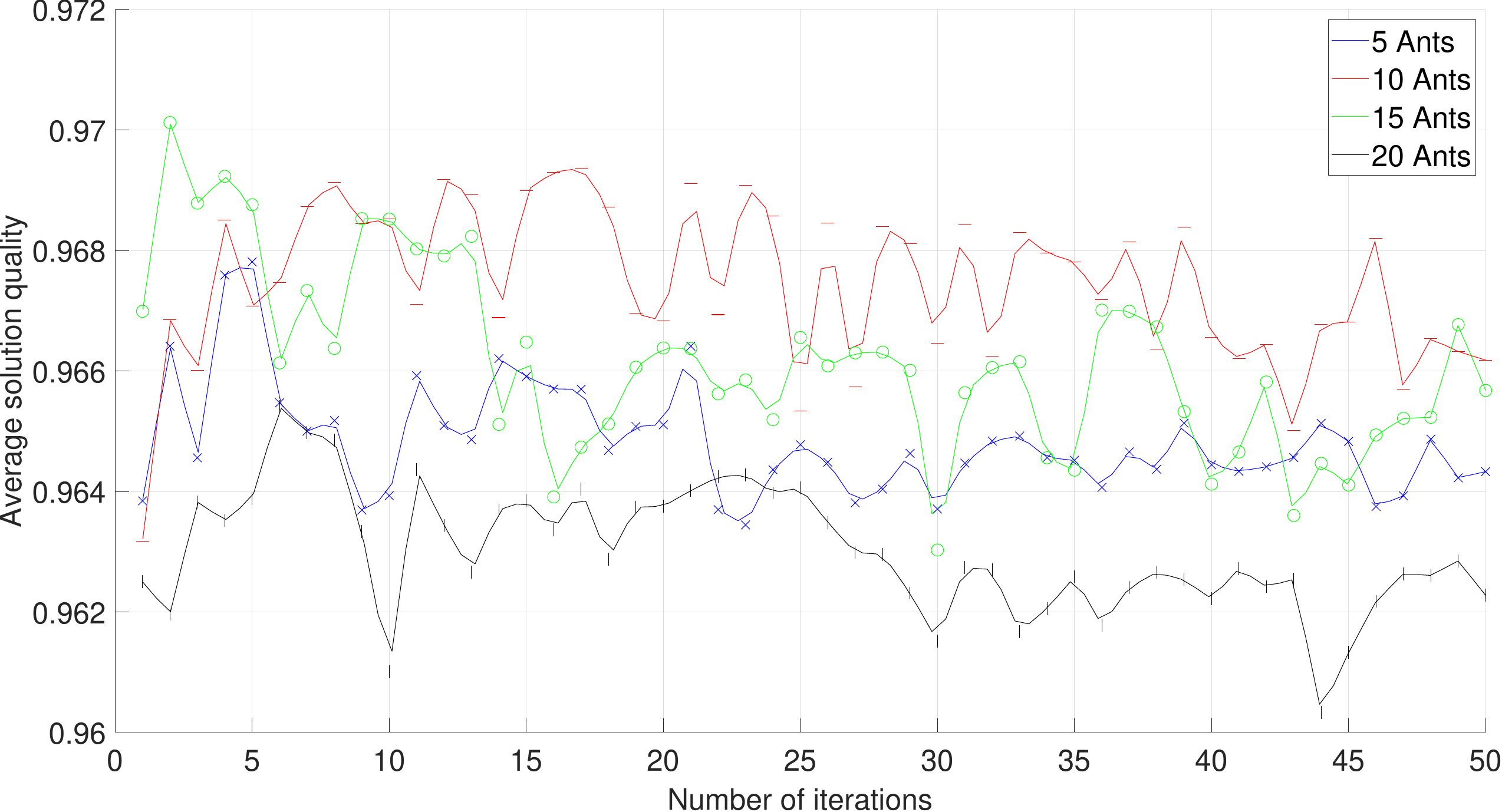}
    \caption{The average grades per iteration with different numbers of ants.}
    \label{fig:ACOResults}
\end{figure}
ACO depends primarily on $\alpha,\beta,\rho$ and the colony size. Following \cite{yu2011ant} we tested $\alpha\in\{1,2,3\}$, $\beta\in\{2,4,6\}$ and $\rho\in\{0.10,0.05,0.01\}$. The combination $\alpha=1.0,\beta=2.0,\rho=0.01$ performed best and was used thereafter. Colony sizes of $5$, $10$, $15$, and $20$ ants were compared (Fig. \ref{fig:ACOResults}), where $10$ ants produced the best average results.

\begin{figure}
    \centering
    \includegraphics[width=0.95\columnwidth]{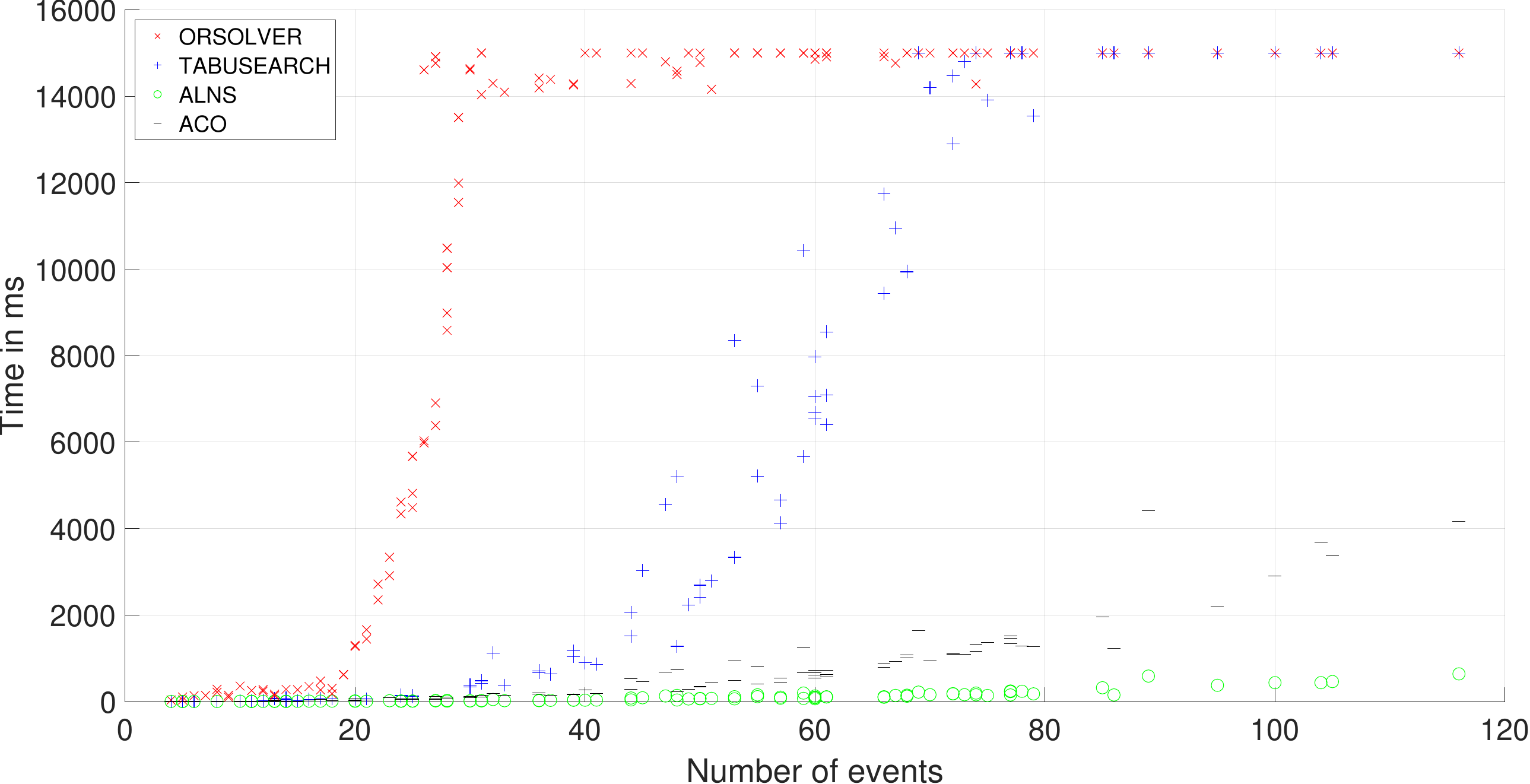}
    \caption{The solution time required by the solution algorithms for different numbers of events.}
    \label{fig:algoResults}
\end{figure}

After determining all parameters of the metaheuristics, all solution methods can be compared with each other. For the average quality, either the exact solution of the solver or the best-known solution of the other solution methods was used. Figure \ref{fig:algoResults} shows the solution times of the four solution methods for different numbers of events. It can be seen that the MIP solver has a strongly increasing runtime from $20$ optimized events onward, up to the limit of $15$ seconds. This $15$-second limit is an arbitrary limit, set by the project settings, at which the user should not wait any longer. This is due to the fact that the system is also used in real-time scenarios in a car. Contrary to the exact solution, the metaheuristics run with an increasing number of events at a low runtime at the expense of the quality of the solutions. Firstly, TS runtime increases notably above about $45$ events while maintaining roughly $99\%$ average quality, followed by ACO with a quality of about $97\%$ and a seemingly constant increase in runtime from $60$ events onward. ALNS has the worst quality with $94.7\%$, which is still very good, and an almost constant runtime, which was able to solve even the largest problem instances in the test series in less than a second.

Based on these observations we propose a hybrid, dynamic procedure that trades off optimality and responsiveness: the MIP solver is used for small problems ($N\in[0,15)$). TS is applied to medium-sized problems (e.g. $N\in[15,45)$), where it offers high quality until runtimes grow. And for larger problems ACO is preferred for its solution quality, with ALNS as an interchangeable option if ACO’s runtime budget becomes insufficient.

\section{Conclusion}\label{sec:Conclusion}
This paper introduces a novel EVRP formulation that integrates multi-day scheduling with EV-specific constraints such as limited battery range and parallel charging. We propose an MIP model that handles both time-fixed and time-flexible events and compare an exact OR-Tools solver with three metaheuristics: TS, ALNS, and ACO. Experiments on randomized instances show that, while the MIP solver produces optimal results for small instances, it becomes impractical as problem size increases. In contrast, the metaheuristics yield near-optimal solutions with substantially shorter runtimes: TS is most effective for medium-sized instances, whereas ACO and ALNS scale better for large instances. Limitations of the current model include the omission of dynamic traffic, nonlinear charging curves, and fluctuating station availability. Future work should investigate hybrid pipelines that apply exact solving to critical subproblems and metaheuristics for large-scale search, and integrate more realistic operational models to enhance real-world applicability.

\vspace{12pt}

\end{document}